# NEMO5: Achieving High-end Internode Communication for Performance Projection beyond Moore's Law

Authors in alphabetical order: Robert Andrawis, Jose David Bermeo, James Charles, Jianbin Fang, Jim Fonseca, Yu He, Gerhard Klimeck, Zhengping Jiang, Tillmann Kubis, Daniel Mejia, Daniel Lemus, Michael Povolotskyi, Santiago Alonso Pérez Rubiano, Prasad Sarangapani, Lang Zeng

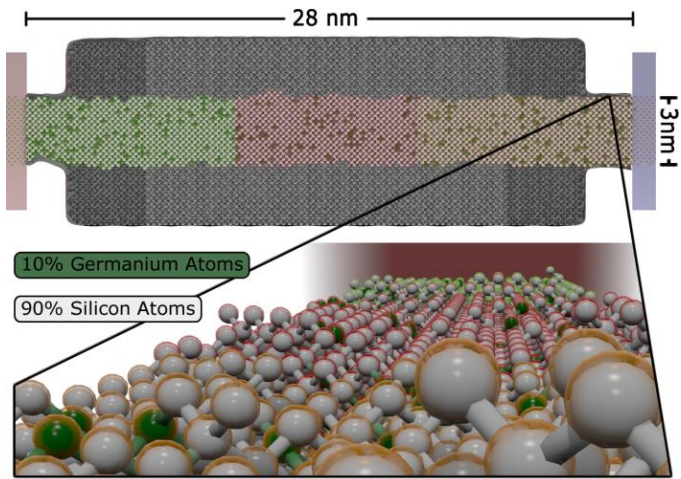

*Figure 1: Atomically resolved UTB device considered in this work. The confinement direction is 3nm long only. Surface roughness is shown in the inset.*

## Abstract

Electronic performance predictions of modern nanotransistors require nonequilibrium Green's functions including incoherent scattering on phonons as well as inclusion of random alloy disorder and surface roughness effects. The solution of all these effects is numerically extremely expensive and has to be done on the world's largest supercomputers due to the large memory requirement and the high performance demands on the communication network between the compute nodes. In this work, it is shown that NEMO5 covers all required physical effects and their combination. Furthermore, it is also shown that NEMO5's implementation of the algorithm scales very well up to about 178176CPUs with a sustained performance of about 857 TFLOPS. Therefore, NEMO5 is ready to simulate future nanotransistors.

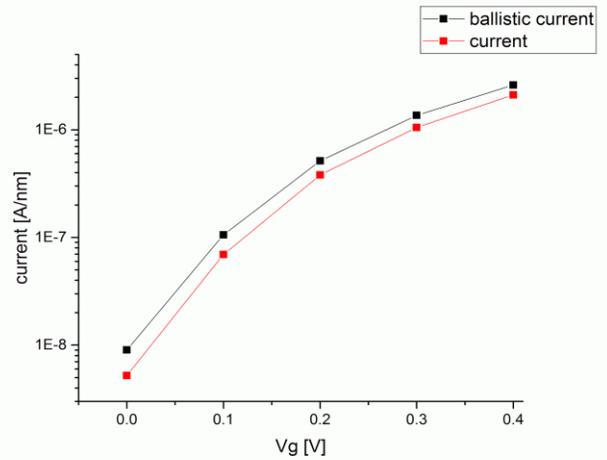

*Figure 2: Current voltage characteristics of the considered UTB device with (red) and without (back) incoherent scattering on optical and acoustic phonons.*

## Overview of the problem and its importance

State of the art and future semiconductor logic devices are in the nanometer length scale. The thickness of ultrathin body (UTB) transistors in the year 2020 is predicted to be around 3nm [1]. Given that a typical distance between two semiconductor atoms is about 0.1nm, the number of atoms is countable. Atom sized device features cannot establish material properties that require a large number of atoms. Therefore, the prediction of future device properties cannot rely on material properties. Instead, detailed calculations of the concrete device structure in subatomic resolution is vital: Device defects and imperfections, such as alloy and dopant atom distributions or roughness fluctuations influence the device performance. For any finite temperature, device atoms vibrate around their ideal lattice position, i.e. they support phonons. Device electrons scatter on these phonons incoherently: This randomizes electron energy and momentum. Such scattering is an important contribution to the device resistance and has to be included in reliable performance predictions as well. These incoherent effects can counterbalance or enhance the coherent nanoscale effects such as tunneling and confinement. For instance, random device fluctuations can confine (localize) electrons in some device sections and delocalize in others. To reliably predict nanoscale device performance requires a consistent treatment of all above effects.

This work considers electronic transport in ultra-thin double gated nanotransistors (Fig. 1). The device is

28 nm long in its transport direction and has a body thickness of 3nm. The device is assumed to be periodic in the in-plane direction perpendicular to transport. Devices with any kind of random disorder are strictly speaking non periodic. Therefore, the simulated devices extend in the periodic direction beyond the minimum ideal unit cell. In fact the simulated device is as large as possible in the periodic direction to avoid artificial effects due to assumed periodicity. All nanodevices in this work are atomically resolved - typically with about 24000 atoms. The structures in this work are all Silicon alloyed with 10% Germanium. This constellation can be modeled with two approaches: The first approach, i.e. the so-called "virtual crystal approximation" (VCA) [2] idealizes a fictitious atom type that has 90% Si and 10% Ge properties. The ideal approach considers Si and Ge atoms explicitly. The Ge atoms are randomly distributed among the simulation domain. The random nature of this method requires solving all observables for many (>100) different Ge distributions (samples) and results are averaged afterwards. Electrons are represented in the sp3d5s* empirical tight binding method, i.e. each atom hosts 10 orbitals [3] which add up to a matrix rank of 240,000. The simulations include scattering on all relevant phonons (inelastic scattering on optical and elastic scattering on acoustic phonons) as well as charge self-consistent solutions of the Poisson equation. NEMO5 covers all these features, but their solution is notoriously expensive and requires massively parallel computer systems (as discussed in later sections). This work shows NEMO5 supports all required physics and scales very well for these type of computational problems.

Figure 2 illustrates NEMO5 covers the impact of phonon scattering on the device current-voltage characteristics a comparison of the ballistic current vs. applied gate voltage to the scattered result including acoustic and optical phonon scattering.

Reference [17] shows NEMO5's unique and most recent method to model randomness in the device and in the leads.

**Quantitative discussion of current state of the art for science and performance**

It is widely accepted that the nonequilibrium Green's function method (NEGF) consistently describes coherent and incoherent transport. Small size effects, such as tunneling, confinement and interferences as well as any sort of incoherent scattering are treated on an equal footing in the NEGF method. However, the NEGF method is numerically very expensive: It requires the solution of 4 nonlinear partial differential equations that are mutually coupled. The mutual dependence of these equations requires to keep solutions of individual equations in memory. Given the large number of variables of each NEGF equation (atomic resolution typically requires about a quarter million variables for electron transport in this work) and given that each equation has to be solved for a large number of parameters (energy and momentum give about 16.000 energy and momentum tuple) and many voltage and randomness configurations, large, massively parallel compute clusters are inevitable for these simulation tasks.

In addition, the prediction of the electron density and its distribution in the nanodevice requires to couple the NEGF method charge self-consistently with the Poisson equation. To predict current-voltage characteristics, the solution of the NEGF and Poisson equations have to be repeated for each voltage-boundary configuration. When randomness is present in the device, such as in the case of disordered alloys or rough interfaces, NEGF/Poisson calculations for many randomness samples are needed to increase the reliability of the observations.

The NEGF method had been applied on a great variety of transport problems, ranging from phononic [4] to electronic transport [5, 6], covering metals [7, 8], semiconductors [9] as well nanotubes [10, 11] and fullerenes [12] or even (organic) molecules [13,14]. The complexity of the NEGF method often motivates approximations such as nanometer-only resolutions, neglect of incoherent scattering, assumption of ideal device fabrication, etc. Due to the immense numerical load, calculations of ITRS relevant, concrete devices in atomic resolution including incoherent scattering are very rare compared to the abundance of NEGF publications. The references [15,16] are such exceptions and are all based on earlier incarnations of the NEMO-NanoElectronic MOdeling tool. NEMO5, the latest version of the NEMO tools had been designed to support all possible nanodevice simulation needs including and exceeding those functionalities of predecessor NEMO versions. NEMO5 contains an important addition to the modeling capabilities of the NEMO tools: It allows for modeling of non-ideal leads which is vital for reliable assessments of the impact of randomness on the device performance [17]. Compared to earlier NEMO versions, NEMO5 handles the NEGF equations numerically more efficiently: By exploiting analytical dependencies of the Green's functions and self-energies, i.e. all solution functions of the NEGF equations, NEMO5 can systematically avoid about 50% of the calculations of earlier NEMO versions while still preserving the full accuracy of all NEGF equations. While this improvement saves a lot CPU hours, it simultaneously reduces the ratio of computation vs. communication. NEMO5 also allows computational support by coprocessors (such as Intel Xeon Phi) and GPUs. NEMO5 is academic-open source and used among many groups in academia and industry.

**Claims made for innovation and its implementation**

This work is an important milestone to reliably assess the relevance of chip-fabrication typical alloy disorder and surface roughnesses for the performance of next transistor generations. In contrast to typical studies, this work combines the randomness with incoherent scattering on all relevant phonons. This way, the balance of coherent and incoherent effects in the presence of randomness is assessable. A conclusive assessment of this balance with a statistics of 100 random samples for one concrete UTB transistor requires about 20 million CPU hours (with typically 100 iterations of the NEGF and Poisson equations per bias point) on at least several thousand nodes due to the high memory usage. Since this numerical load exceeds the computational infrastructure currently available to the project's team, this work focuses on important algorithmic improvements and efficient numerical implementations in NEMO5. This work shows that 1) all involved physics are covered and 2) the algorithm implementation of NEMO5 utilizes supercomputing performance to a very high level of efficiency.

All electronic transport properties are solved with the NEGF method. In the stationary limit, the NEGF equations are solved with the electronic "lesser than" and retarded Green's functions, $G^<$ and $G^R$, respectively. Their differential equations read in matrix form

$$G^R(k,E) = (E - H(k) - e\Phi - \Sigma^R(k,E))^{-1}, \quad (1)$$
$$G^<(k,E) = G^R(k,E)\Sigma^<(k,E)G^{R\dagger}(k,E). \quad (2)$$

All Green's functions and scattering self-energies $\Sigma^<$ and $\Sigma^R$ are functions of the electronic energy E and transverse momentum k. Scattering is represented with acoustic and optical deformation potential phonons. In this case, the total self-energies read

$$\Sigma^{R,<}(k,E) = \Sigma^{R,<}_{lead}(k,E) + \Sigma^{R,<}_{optical}(k,E) + \Sigma^{R,<}_{acoustic}(k,E), \quad (3)$$

$$\Sigma^{R,<}_{acoustic}(k,E) = \frac{D^2 k_B T}{2\hbar\omega_D \rho v^2 A} \int dk'\, diag G^{R,<}(k',E), \quad (4)$$

$$\Sigma^<_{optical}(k,E) = \frac{\hbar \Xi^2}{8\pi^2 \omega_0 A} \int dk' [(1+n_0)\, diag G^<(k', E+\hbar\omega_0/2\pi) + n_0\, diag G^<(k', E-\hbar\omega_0/2\pi)], \quad (5)$$

$$\Sigma^R_{optical}(k,E) = \frac{\hbar \Xi^2}{8\pi^2 \omega_0 A} \int dk'[(1+n_0)\, diag G^R(k', E-\hbar\omega_0/2\pi) + n_0\, diag G^R(k', E+\hbar\omega_0/2\pi), +0.5\, diag G^<(k', E-\hbar\omega_0/2\pi) - 0.5\, diag G^<(k', E+\hbar\omega_0/2\pi)] \quad (6)$$

It is worth to mention Eq.(6) contains in its exact representation a principal value integral that is usually ignored due to its small contribution to the resonant energies only [15,18,19]. Here, the equations are given in their actual implemented shape. All integrals run over all electronic momenta in the first Brillouin zone. In the Eqs.(4)-(6), the deformation potential of acoustic phonons $D$, the sound velocity $v$, the Debye frequency $\omega_D$, the optical phonon frequency $\omega$, the lattice constant of Silicon a, and the deformation potential of optical phonons $\Xi$ are taken from experimental publications [20]. The Planck constant $h$, the Boltzmann constant $k_B$ and the area covered by a single atom perpendicular to the periodic device direction $A$ are given by nature. The temperature T agrees with room temperature throughout this work. The self-energies $\Sigma^{R,<}_{lead}$ in Eq.(3) describe the coupling of device electrons with the charge reservoirs via 2 semi-infinite leads. In the case of a VCA description of the device alloy, the leads are solved with the transfer matrix method of Ref. [21], whereas in the case of discrete random Ge distributions, the leads include the randomness as well and are therefore solved with NEMO5's adaption of the complex absorbing potential method [17]. Due to their mutual dependence, the Eqs.(1) - (6) are solved iteratively. Once converged, observables such as density n and charge current density j can be solved

$$n = \int dE\, n(E) = \frac{1}{(2\pi)^2} \int dE \int dk\, Im[diag\, G^<(k,E)], \quad (7)$$

$$j = \int dE\, j(E) = \frac{e}{h\pi} \int dE \int dk\, diag[H(k)G^<(k,E) - G^<(k,E)H(k)]. \quad (8)$$

The density of Eq.(7) enters the Poisson equation which is solved for the electrostatic potential Φ

$$\nabla \varepsilon \nabla \Phi = e\,(n - N_D), \quad (9)$$

with ε, the dielectric constant, $e$ the electron charge and $N_D$ the doping density. Since the electrostatic potential enters Eq.(1), all NEGF equations need to get solved with Eq.(9) iteratively until convergence.

It is important to note that NEMO5's formulation of the retarded scattering self-energy in Eqs.(4) and (6) does not require knowledge of any "greater than" functions (such as $G^>$ or $\Sigma^>$). While formulations that require "greater than" functions are physically equivalent, they solve more NEGF equations and keep more Green's functions in memory. Earlier massively parallel NEMO versions solved the retarded Green's functions multiple times on different nodes to avoid communication of large data sets. Compared to that, NEMO5's NEGF algorithm requires about 50% fewer operations.

The energy E and momentum k parameters of the NEGF equations are parallelized with MPI. Each MPI-rank either utilizes multiple OpenMP threads on the CPU-based host or an Intel Xeon Phi coprocessor to parallelize the matrix operations for each individual (E,k) tuple. Communication between different MPI processes is required to solve the integrals in Eqs.(4)-(9). In particular the integrals in Eqs.(4)-(6) are iteratively solved with Eqs. (1) and (2). Depending on the number and concrete values of the (E,k) tuple per of each MPI rank, the amount of communication and local contribution to each integral in the Eqs. (4)-(6) varies significantly. To avoid idling ranks, all possible local calculations and integrations are done before the blocking MPI communication of the Green's functions. To avoid deadlocks, the communication tasks are ordered prior to the calculation of Eqs. (4)-(6). To avoid the complex graph coloring that a sorting algorithm typically entails, NEMO5 determines the communication task order with respect to the (E,k) values "on-the fly", i.e. during the solution of Eqs.(4)-(6). In this scheme, however, great care is put into the distribution of (E,k) points among the available MPI-ranks - prior to any NEGF solution.

**Specific application(s) used to measure performance**

NEMO5's custom internal tic-toc profiling system measures time and memory consumption to provide rank-specific data on selected blocks of code. The low-overhead system allows for hierarchical profiling data as defined by code developers, which is then written in XML format and can be displayed on an interactive webpage for visualization.

The number of floating point operations was measured with Intel vTune Amplifier XE 2013 update 15 and resulted in $2.866 \times 10^{17}$ operations. This resulted on 178176 CPUs of Tianhe-2 in 857 TFLOPS (floating point operations per second). Ganglia and the Unix tool top was used to measure peak memory usage.

**System and environment where performance was measured, including specific measurement methodology**

Scaling tests for the simulation of interest were run on multiple supercomputer systems: Conte at RCAC, Purdue University, TACC Stampede at the University of Texas at Austin, and Tianhe-2 at the National Supercomputer Center in Guangzhou, China.

Although the devices modeled for these scaling tests always represent the same physical device, the sizes of the matrices for which computations take place were modified to meet memory and time constraints. By choosing a large value for the device's periodic dimension, the periodic distribution of random alloy atoms is repeated less frequently, making the transport model more realistic (see discussion above). Because of this, a UTB device that is "thicker" along its periodic dimension is encouraged. The limitations to this advantage are memory constraints and increased time to solution due to larger matrix dimensions.

Due to the diverse capabilities of supercomputers available to solve these simulations, three uniform device "thicknesses" were used for strong and weak scaling tests: device cross-sections in crystal lattice unit cells of 6x10, 6x6, and 6x2 unit cells. 6x10, 6x6 and 6x2 unit cell cross section devices contain sparse and dense matrices with $1200^2$, $720^2$ and $240^2$ elements, respectively.

Due to the nature of the NEGF algorithm and Eqs. (4) – (6), which involve integration of the diagonals of Green's functions, the solution of each energy-momentum (E-k) tuple increases the memory footprint. For example, a 6x10 unit cell cross section device contributes about 4.5 GB memory per E-k tuple solved in series on each MPI process. If two MPI processes are present on each node, the memory footprint increases in steps of about 9 GB for each E-k tuple solved in series. If four E-k tuples are solved in series on two MPI processes, the system's peak memory is about 53 GB for a 6x10 unit cell device.

On Conte and Tianhe-2, NEMO5 was compiled and run with Intel Composer XE 2013 SP1 update 2, with Intel C++ compiler version 14.0.2, MKL version 11.1.2 and Intel MPI version 4.1.1. On Stampede, NEMO5 was compiled and run with Intel Composer XE 2013 SP1 update 1, with Intel C++ compiler version 14.0.1, MKL version 11.1.1 and Intel MPI version 4.1.3. The Intel Xeon Phi coprocessors available on all above listed machines can be used for the NEGF transport problem in NEMO5. In NEMO5, the Intel Xeon Phi coprocessors are used in the compiler assisted offloading mode of Intel MKL dense matrix routines. In particular, matrix-matrix multiplications (zgemm) and matrix inversions (zgesv) perform very well on the coprocessors if the matrix size exceeds about 2000 rows and columns.

Conte is hosted at Purdue's Rosen Center for Advanced Computing (RCAC) and has 580 nodes, 16 cores each with 10336 Intel Xeon-E5 cores total configured with 64GB RAM. Each node has two 60 core Intel Xeon Phi cards. 74 nodes are dedicated to nano simulations.

Tianhe-2, located in Guangzhou, China, has 16,000 24-core nodes, each with 64 GB of memory, and three Intel Xeon Phi Coprocessors per node.

The Stampede supercomputer at the University of Texas at Austin has 6400 16-core nodes, most of which have an Intel Xeon Phi Coprocessor, and 32 GB of memory per node.

All scaling data show the performance of one solution of the Eqs.(1)-(6).

**Performance results, including scalability (weak and strong), time to solution, efficiency (of bottleneck resources), and peak performance**

All strong scaling results are given in Fig. 3. Due to Conte's small number of nodes but large available memory, the best type of scaling test to perform on Conte for the device of interest was strong scaling. A 6x10 unit cell cross section with $1200^2$ element matrices was used for this model. For a host-only strong scaling test on Conte, three distributions of 720 E-k tuples were simulated. For each strong scaling data point, 2 MPI processes were executed on each node and used multithreading provided by Intel MKL routines so that each would use 8 cores for dense matrix computations. The scaling data points involved distribution of work on 90, 180 and 360 nodes with 4, 2 and 1 E-k tuples solved per process, respectively. On Conte, aside from the host-only scaling tests, some strong scaling data was gathered with the use of Intel Xeon Phi Coprocessors. For these tests, a single Xeon Phi Coprocessor is used per node by one of the two MPI processes, with each coprocessor using 240 threads for compiler assisted offloaded MKL dense matrix multiplication (zgemm) and dense matrix inversion (zgesv) routines.

Previous tests have shown that compiler assisted offloading performs best with similar systems containing matrices with over $2000^2$ elements. Unfortunately, due to memory limitations of the algorithm of interest, $1200^2$ or less elements must be used, which results in the Xeon Phi and host performance being comparable. This is due to the overhead of communication between the host and Xeon Phi cards before and after matrix computation.

All weak scaling results are shown in Fig. 4. On Stampede a 6x10 unit cell cross section device was tested for weak scaling with Intel Xeon Phi coprocessors included. The reason for weak scaling being tested instead of strong scaling on Stampede is the 32 GB memory limitation, which only allows for two parallel E-k instances of this device to be modeled at a time per node.

On Tianhe-2, due to walltime limitations of 15 minutes, weak scaling of a 6x6 unit cell cross section device was measured.

In general, the scaling is better the larger the system size and the matrix dimensions are. It can be seen from Figs 3 and 4 that all scaling results on Tianhe-2 are superior to the results on the other machines. This is due to a different simulation setting: All energy meshes on conte and Stampede simulations are adaptive and inhomogeneous, while those on Tianhe-2 are homogeneous. The inhomogeneous meshes cause a load imbalance in the current NEMO5 implementation of Eqs.(5) and (6). In contrast, the homogeneous mesh yields a very balanced load on all ranks. The NEMO5 team is correcting the load imbalance in the inhomogeneous case at the time of the submission deadline.

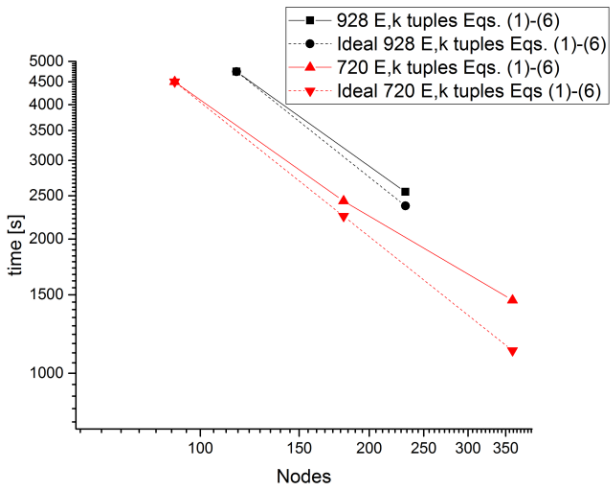

*Figure 3: Strong Scaling on Conte at RCAC, Purdue University. 1 node contains 16 CPU cores.*

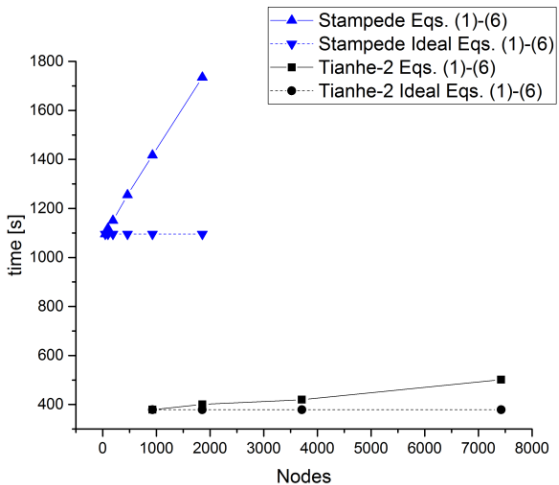

*Figure 4: Weak Scaling on Stampede at TACC, University of Texas at Austin and Tianhe-2 at the National Supercomputer Center in Guangzhou, China. 1 node contains 16 CPU cores on Stampede and 24 CPU cores on Tianhe-2.*

**Implications for future systems and applications**

This work has focused on the realistic modeling of alloyed UTBs with applications to the ITRS 2020 node. The IV curve presented shows the impact of incoherent scattering in the considered devices. Due to the stochastic nature of including explicit atoms simulations considering random alloys and roughness have to be done 100s of times to get a distribution of observables from the random distribution of the atoms in the alloy. In addition to the need for many samples, the simulated device needs to be very large in the periodic direction to avoid artificial effects from repeating random alloy arrays. The inclusion of random alloys on ballistic Si and SiGe wires and the comparison to VCA has been done in ref. [17]. The paper discusses the differences between VCA and random alloy included both in the leads and the device. In the on-state the difference in current can be as much as 45% and is not negligible for accurate modeling of these devices. Similar effects are expected for UTBs and will be investigated in the future.

The numerical performance data convince that NEMO5 is ready to simulate the UTB device including scattering on phonons and random alloy and roughness disorder. The scaling data show that the algorithm runs very efficiently on the supercomputers essential for the simulation load up to about 180000CPUs. Note that NEMO5 requires only 50% of the numerical operations of earlier, massively parallel NEGF implementations.

**References**

1. International Technology Roadmap for Semiconductors [https://www.itrs.net]
2. L. Bellaiche, and D. Vanderbilt, Phys. Rev. B 61, 7877 (2000)
3. T. Boykin, G. Klimeck and F. Oyafuso, Phys. Rev. B 69, 115201 (2004)
4. Y. Xu, J.-S. Wang, W. Duan, B.-L. Gu, and B. Li, Phys. Rev. B 78, 224303 (2008)
5. M. Luisier, A. Schenk, and W. Fichtner, J. Appl. Phys. 100, 043713 (2006)
6. V. N. Do, P. Dollfus, and V. L. Nguyen, J. Appl. Phys. 100, 093705 (2006)
7. Z. Chen, J. Wang, B. Wang, and D. Y. Xing, Phys. Lett. A 334, 436 (2005)
8. Y. Ke, K. Xia, and H. Guo, Phys. Rev. Lett. 100, 166805 (2008)

9. A. Bulusu and D. G. Walker, Transaction of the ASME 129, 492 (2007)
10. T. Yamamoto and K. Watanabe, Phys. Rev. Lett. 96, 255503 (2006)
11. M. Lazzeri, S. Piscanec, F. Mauri, A. C. Ferrari, and J. Robertson, Phys. Rev. Lett. 95, 236802 (2005)
12. H. Li and X. Q. Zhang, Phys. Lett. A 372, 4294 (2008)
13. T. Sato, K. Shizu, T. Kuga, K. Tanaka, and H. Kaji, Chem. Phys. Lett. 458, 152 (2008)
14. P. Damle, A. W. Ghosh, and S. Datta, Chem. Phys. 281, 171 (2002)
15. R. Lake, G. Klimeck, R.C. Bowen, and D. Jovanovic, J. Appl. Phys. 81, 7845 (1997)
16. M. Luisier and G. Klimeck, Phys. Rev. B 69, 115201 (2009)
17. Y. He, Y. Wang, G. Klimeck, and T Kubis, Appl. Phys. Lett. 105, 213502 (2014)
18. S.-C. Lee and A. Wacker, Phys. Rev. B 66, 245314 (2002)
19. N. Vukmirovic´, Z. Ikonic´, D. Indjin, and P. Harrison, Phys. Rev. B 76, 245313 (2007)
20. Springer Materials—The Landolt-Börnstein Database
21. M. Luisier, A. Schenk, W. Fichtner, and G. Klimeck, Phys. Rev. B 74, 205323 (2006)